# Narrow linewidth Brillouin laser based on chalcogenide photonic chip

Irina V. Kabakova,[1-2*] Ravi Pant,[1-2] Duk-Yong Choi,[2-3] Sukhanta Debbarma,[2-3] Barry Luther-Davies,[2-3], Stephen J. Madden,[2-3] and Benjamin J. Eggleton[1-2]

[1]Institute of Photonics and Optical Sciences (IPOS), School of Physics, University of Sydney, NSW, 2006
[2]Centre for Ultrahigh-bandwidth devices for Optical Systems (CUDOS), Australia
[3]Laser Physics Centre, Australian National University, Canberra, ACT, 0200, Australia
*Corresponding author: kabakova@physics.usyd.edu.au

We present the first demonstration of a narrow linewidth, waveguide-based Brillouin laser which is enabled by large Brillouin gain of a chalcogenide chip. The waveguides are equipped with vertical tapers for low loss coupling. Due to optical feedback for the Stokes wave, the lasing threshold is reduced to 360 mW, which is 5 times lower than the calculated single-pass Brillouin threshold for the same waveguide. The slope efficiency of the laser is found to be 30% and the linewidth of 100 kHz is measured using a self-heterodyne method.

Highly coherent, tunable lasers are essential for applications in coherent optical communications [1], remote sensing [2] and microwave photonics [3]. One possible way to build such a laser is to exploit the intrinsically narrow linewidth and tunability of stimulated Brillouin scattering (SBS). Recent studies have demonstrated that Brillouin lasers (BL) can have orders of magnitude higher coherence than their pump sources [4-7]. The phase noise of the pump is transferred to the Stokes wave but with significantly reduced magnitude due to combined action of acoustic damping and cavity feedback [4].

Several geometries for the BL have been demonstrated using silica [4, 5] and highly-nonlinear (chalcogenide) optical fibers [6, 7]. The typical length of the gain fiber in these experiments was tens of meters when using SMF-28 or a couple of meters for a highly-nonlinear fiber. The current focus on integration and miniaturization suggests that a broadly tunable chip-based BL could provide an appealing alternative to existing electrical microwave oscillators and semiconductor lasers. Miniature chip-based solutions require large SBS gain, which can be obtained via resonant enhancement in an optical cavity [8] or in a material with large SBS coefficient such as chalcogenide glass [9-11]. We previously reported cavity enhanced cascaded SBS in a chalcogenide chip by exploiting the end-face feedback with a corresponding reduction in SBS threshold by a factor of 4 [12].

In this paper we present the first demonstration of a waveguide-based, narrow linewidth Brillouin ring laser using a chalcogenide chip. Via the SBS process, the Stokes signal was generated in an $As_2S_3$ waveguide [9] and circulated in a ring fiber cavity comprising the chip (Fig. 1(a)). The pump, however, was removed from the cavity after a single pass. This allowed us to avoid the requirement for precise matching of the cavity free spectral range with the Brillouin frequency shift, necessary for high-Q resonators [8]. In the laser characterization we found that the power threshold and the slope efficiency were 360 mW and 30%, respectively. The laser linewidth was 15 times smaller than the linewidth of the pump and over 300 times narrower than the Brillouin gain bandwidth [9].

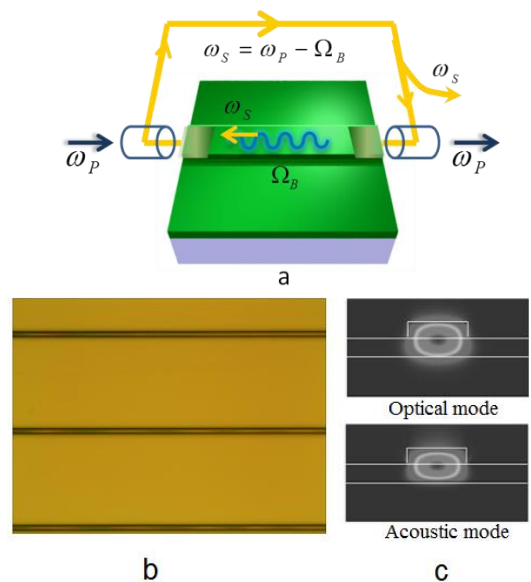

Fig. 1. (a) Schematic of a Brillouin laser based on photonic chip. (b) Micrograph of the $As_2S_3$ waveguides (top view) with vertical tapers [13]. (c) Numerically calculated acoustic and optical modes in the waveguide.

The laser gain medium was a 7 cm-long, 4 µm-wide $As_2S_3$ rib waveguide with the effective area of 2.3 µm² [9-10]. At both ends the waveguide was equipped with vertical tapers which transferred the power adiabatically to 3x3µm SU8 waveguides to achieve low loss coupling to UHNA-4 fiber [13]. The gain coefficient of an $As_2S_3$ waveguide was $g_B = 0.7 \cdot 10^{-9}$ m/W ($g_B \sim 100 \times$ silica) and the Brillouin frequency shift was $\Omega_B \approx 7.5 \pm 0.2$ GHz [9].

Via the SBS process the pump light ($\omega_p$) interacts coherently with the acoustic phonons, leading to inelastic scattering of the pump (Fig. 1(a)). The scattered photons, i.e. Stokes photons ($\omega_s$), are downshifted with respect to the pump by the acoustic phonon frequency ($\Omega_B = \omega_P - \omega_s$). This is determined by the acoustic mode which has the largest overlap with the optical mode [14]. The mode profiles, calculated using COMSOL are shown in Fig. 1(c).

Due to the feedback of the Stokes signal, the optical field at $\omega_s$ builds up inside the waveguide. Lasing occurs when the gain at the Stokes frequency exceeds the cavity loss. The cavity mode, nearest to the peak of the Brillouin gain (free spectral range of the waveguide/fiber loop cavity is FSR=8.4 MHz and the gain bandwidth is $\Delta \nu_B = 34$ MHz), experiences maximum amplification and hence, it lases preferentially.

The experimental setup of the Brillouin ring-cavity laser is shown in Fig. 2. The pump from a continuous wave (CW) tunable laser with a bandwidth of $\Delta f_P = 1.55$ MHz was amplified using an erbium-doped fiber amplifier (EDFA) before being directed to the circulator. A 99/1% coupler (C1) was inserted to monitor the input power to the waveguide with a power meter (PM). From port (2) of the circulator the pump light was butt-coupled to the photonic chip using UHNA-4 silica fibers and refractive index matching liquid.

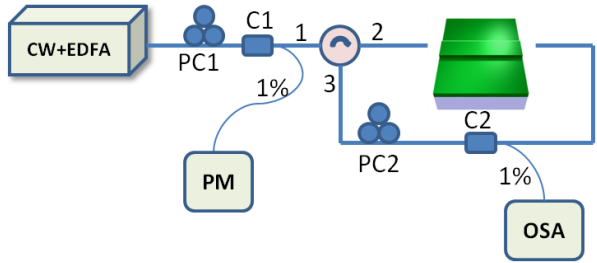

Fig. 2. Experimental setup.

The pump made a single transit of the cavity before being stripped out by the circulator. The Stokes signal generated in the waveguide, however, was redirected to port (3) and oscillated inside the cavity. The Stokes wavelength could be aligned to the cavity resonance by tuning the pump wavelength.

The number of round trips for the Stokes wave in the ring cavity was limited by the optical losses. These included: (a) propagation loss in the waveguide (-1.4dB), (b) the coupling loss (due to vertical tapers this was reduced from –4dB [9] to –2dB per facet) and (c) losses in other fiber-optic elements (-1dB in total). The polarization of the pump and the Stokes signal was controlled using fiber polarization controllers PC1 and PC2, respectively. The 1% port of the coupler (C2) was connected to an optical spectrum analyzer (OSA) and served to monitor the laser output.

The power in the Stokes signal versus the coupled pump power is plotted in Fig. 3. Just above a pump power of $P_L$=0.36 W, the Stokes power increased by 37 dB, from -30 dBm to +7 dBm indicating that the lasing threshold was near $P_L$. The slope efficiency was estimated to be 30%.

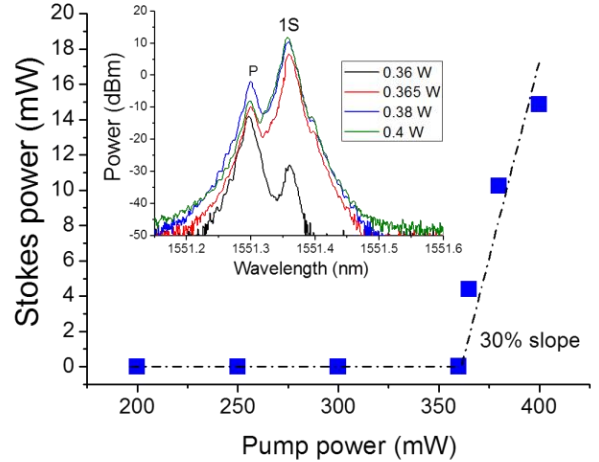

Fig. 3: The Stokes power versus the input pump power (markers). The inset shows back-scattered OSA spectra taken from 1% port of the C2 coupler. The two signals are the pump (P) and the Stokes (S), which grows dramatically for input coupled powers >0.36 W.

The inset in Fig. 3 shows spectra of the backscattered light circulating in the ring-cavity, measured using the OSA. The different colored lines correspond to the back-scattered signals for varying pump power close to the lasing threshold. Two wavelengths can be identified: the pump signal at $\lambda_P = 1551.3$ nm and the generated Stokes signal at $\lambda_S = 1551.94$ nm (the Stokes shift in chalcogenide is $\Delta \lambda_B^{ch} = 0.064$ nm, compared to $\Delta \lambda_B^{silica} = 0.08$ nm in silica fiber). The signal at $\lambda_P$ comes from the residual back-reflection of the pump from the front facet of the chip. It is, however, much weaker than the coupled pump power (by more than -20dB) and the Stokes wave (above the lasing threshold). It can be removed completely using a band-pass filter or a fiber Bragg grating.

Without the cavity feedback the threshold for generation of the Stokes signal in a 7 cm long $As_2S_3$ waveguide was calculated to be 1.73 W, i.e. five times higher than that with the feedback [15]. Fig. 4 shows back-scattered light from the chip when the fiber loop was disconnected and no feedback for the Stokes wave existed. For the same amount of pump power (P=0.4 W), the Stokes signal is hardly noticeable, in contrast to the green curve in the inset to Fig. 3.

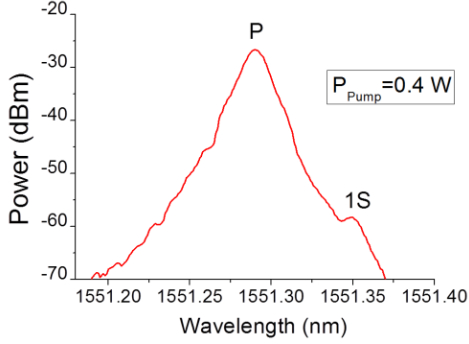

Fig. 4: The back-scattered signal for the maximum coupled peak power (green curve in the inset of Fig. 3), but with the broken feedback around the chip.

To measure the linewidth of the BL we applied a self-heterodyne technique [16] (a schematic of the setup is shown in Fig. 5). A fiber Bragg grating (FBG) was first used to remove the pump from the laser output. The resulting Stokes signal was then sent to a Mach-Zehnder (MZ) interferometer. The signal in one arm was modulated using an intensity modulator (IM) driven by a radio frequency generator (RFG) to create sidebands, whilst the signal in the other arm was delayed in a 40 km-long standard silica fiber to break the coherence between the two arms. By adjusting the bias applied to the IM we could suppress the main carrier almost completely. However, the bias drifted somewhat over the course of the experiment resulting in on-average a non-zero component at $\lambda_s$. After the MZ interferometer the two outputs were combined and sent to a photo detector (PD) connected to a radio frequency spectrum analyzer (RFA) Agilent E4448A with the maximum spectral resolution of 3 Hz. The width of the RF spectrum was used to determine the linewidth of either the Stokes signal or the pump laser.

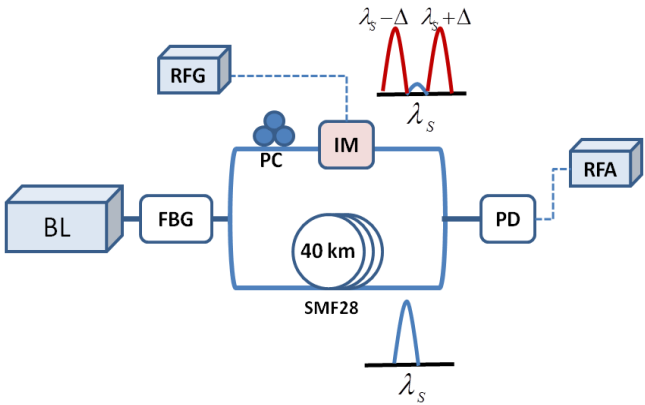

Fig. 5. Schematic of the measurement setup used to obtain the Brillouin laser (BL) linewidth. Notation used in the scheme: FBG-fiber Bragg grating, PC-polarization controller, IM- intensity modulator, RFG-radio frequency generator, PD- photo detector, RFA-radio frequency analyzer.

Fig. 6 shows the result of this measurement: the blue line corresponds to the pump, whereas the red curve is the measurement of the lasing signal. The 3dB bandwidth of Stokes signal was ≈ 100 kHz, i.e. 15 times smaller than that of the pump where $\Delta f_P = 1.55$ MHz. A small delta function-like peak on the top of Lorentzian pedestal for the Stokes measurement results from beating of the residual coherent components, which is caused by imperfect suppression of the main carrier in the Mach-Zehnder. It is, therefore, a measurement artifact and the Stokes linewidth should be calculated as the full width of half maximum of the Lorentzian pedestal.

The effect of line narrowing in Brillouin lasers has been studied theoretically and experimentally [4-5]. The full width at half maximum (FWHM) of the Stokes signal $\Delta f_S$ can be approximated by $\Delta f_S = \Delta f_P/K^2$, with $\Delta f_P$ being the pump FWHM and $K = 1 + \gamma_a/\Gamma_c$. Here $\gamma_a = 2\pi\Delta\nu_B$ is the damping rate of the acoustic wave ($\Delta\nu_B = 34$ MHz [8]) and $\Gamma_c = -c \ln R^2/nL$ is the loss rate of a cavity with the feedback $R$, length $L$ and the refractive index of the medium $n$. If the cavity loss is low, $K > 1$ and $\Delta f_S < \Delta f_p$. Thus, BL can have higher coherence than the pump laser and, hence, this technique can be used to purify the pump source.

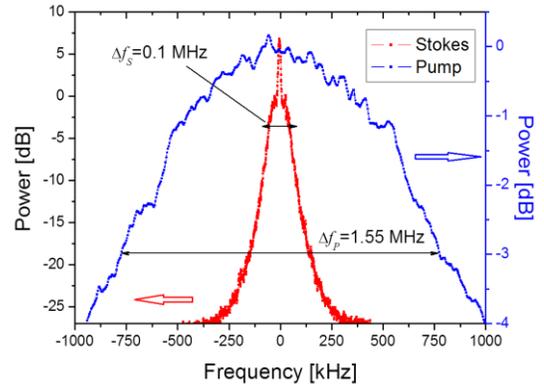

Fig. 6. The FWHM of the Brillouin laser measured using RF analyzer. The linewidth is measured at a 3 dB level of a Stokes pedestal.

In our experiment the ring cavity consisted of a silica fiber (10 m) and a chalcogenide waveguide (7 cm long), i.e. the optical length was $L \cdot n = 14.67$ m (1.45x10m + 2.45x0.07m). Based on the coupling and propagation losses of the waveguide and optical components in the fiber loop, we calculated the feedback parameter to be R=14.4%. This allows us to find a theoretical estimate for the line narrowing factor $K^{th} = 3.38$ and $\Delta f_S^{th} = \Delta f_P /11.43$ which compares favorably with the measured value of $K^{exp} = 3.93$. The small discrepancy could arise from a slightly wrong estimate of the cavity length or the cavity loss. We note that the laser linewidth is over 300 times narrower than the Brillouin gain bandwidth in chalcogenide glass.

In conclusion, we have demonstrated a 100 kHz linewidth Brillouin laser based on chalcogenide photonic chip. Due to optical feedback inside the ring-cavity, the lasing threshold is at least 5 times lower than the single pass Brillouin threshold in the same waveguide. In addition, the linewidth of the laser is measured to be 15 times narrower than the pump linewidth and 300 times narrower than the Brillouin gain bandwidth. The narrowing factor $K$ is limited by the coupling and propagation losses in our waveguides, which can potentially be improved in the future and which could lead to sub-kilohertz spectral linewidth. Ideally, the couplers and the ring cavity would need to be integrated on a chip together with the gain waveguide. This would substantially increase the laser stability and reduce further the lasing threshold. The work is this direction is currently in progress [17].

### Acknowledgements
Funding from the Australian Research Council (ARC) through its Laureate Project FL120100029 is gratefully acknowledged. This research was also supported by the ARC Center of Excellence for Ultrahigh bandwidth Devices for Optical Systems (project number CE110001018).

### References
[1] E. Ip, A. P. T. Lau, D. J. F. Barros, J. M. Kahn, Opt. Express **16**, 753 (2008).
[2] C. Karlsson, F. Olsson, D. Letalick, and M. Harris, Appl. Opt. **39**, 3716 (2000).
[3] T. M. Fortier, M. S. Kirchner, F. Quinlan, J. Taylor, J. C. Bergquist, T. Rosenband, N. Lemke, A. Ludlow, Y. Jiang, C. W. Oates and S. A. Diddams, Nat. Photonics **5**, 425 (2011).
[4] A. Debut, S. Randoux, and J. Zemmouri, Phys. Rev. A **62**, 023803 (2000).
[5] A. Debut, S. Randoux, and J. Zemmouri, J. Opt. Soc. Am. B **18**, 556 (2001).
[6] K. H. Tow, Y. Leguillon, S. Fresnel, P. Besnard, L. Brilland, D. Mechin, D. Tregoat, J. Troles, and P. Toupin, Opt. Express **20**, B104 (2012).
[7] K. H. Tow, Y. Leguillon, P. Besnard, L. Brilland, J. Troles, P. Toupin, D. Mechin, and S. Molin, Optics Lett. **37**, 1157 (2012).
[8] H. Lee, T. Chen, Jiang Li, K. Y. Yang, S. Jeon, O. Painter and K. J. Vahala, Nat. Photon. **6**, 369 (2012).
[9] R. Pant, C. G. Poulton, D.-Y. Choi, H. Mcfarlane, S. Hile, E. Li, L. Thevenaz, B. Luther-Davies, S. J. Madden, and B. J. Eggleton, Opt. Express **19**, 8285 (2011).
[10] R. Pant, A. Byrnes, C. G. Poulton, E. Li, D.-Y. Choi, S. Madden, B. Luther-Davies, and B. J. Eggleton, Optics Lett. **37**, 969 (2012).
[11] S. Levy, V. Lyubin, M. Klebanov, J. Scheuer and A. Zadok, Optics Lett. **37**, 5112 (2012).
[12] R. Pant, E. Li, D.-Y. Choi, C. G. Poulton, S. J. Madden, B. Luther-Davies, and B. J. Eggleton, Opt. Lett. **36**, 3687-3689 (2011).
[13] S. Madden, Z. Jin, D. Choi, S. Debbarma, D. Bulla, and B. Luther-Davies, Opt. Express **21**, 3582 (2013).
[14] P. T. Rakich, C. Reinke, R. Camacho, P. Davids, and Z. Wang, Physical Review X **2**, 011008 (2012).
[15] S. Norcia, S. Tonda-Goldstein, D. Dolfi, J.-P. Huignard, and R. Frey, Optics Lett. **28**, 1888 (2003).
[16] L. E. Richter, H. I. Mandelberg, M. S. Kruger, and P. A. McGrath, IEEE J. Q. Electron. **22**, 2070 (1986).
[17] H. G. Winful, I. V. Kabakova, and B. J. Eggleton, Opt. Express **21**, 16191 (2013).

## References

[1] E. Ip, A. P. T. Lau, D. J. F. Barros, J. M. Kahn, "Coherent detection in optical fiber systems," Opt. Express **16**, 753 (2008).

[2] C. Karlsson, F. Olsson, D. Letalick, and M. Harris, "All-Fiber Multifunction Continuous-Wave Coherent Laser Radar at 1.55 μm for Range, Speed, Vibration, and Wind Measurements" Appl. Opt. **39**, 3716 (2000).

[3] T. M. Fortier, M. S. Kirchner, F. Quinlan, J. Taylor, J. C. Bergquist, T. Rosenband, N. Lemke, A. Ludlow, Y. Jiang, C. W. Oates and S. A. Diddams, "Generation of ultrastable microwaves via optical frequency division," Nat. Photonics **5**, 425 (2011).

[4] A. Debut, S. Randoux, and J. Zemmouri, "Linewidth narrowing in Brillouin Lasers: Theoretical analysis," Phys. Rev. A **62**, 023803 (2000).

[5] A. Debut, S. Randoux, and J. Zemmouri, "Experimental and theoretical study of linewidth narrowing in Brillouin fiber ring lasers," J. Opt. Soc. Am. B **18**, 556 (2001).

[6] K. H. Tow, Y. Leguillon, S. Fresnel, P. Besnard, L. Brilland, D. Mechin, D. Tregoat, J. Troles, and P. Toupin, "Linewidth-narrowing and intensity noise reduction of the 2nd order Stokes component of a low threshold Brillouin laser made of $Ge_{10}As_{24}Se_{68}$ chalcogenide fiber," Opt. Express **20**, B104 (2012).

[7] K. H. Tow, Y. Leguillon, P. Besnard, L. Brilland, J. Troles, P. Toupin, D. Mechin, and S. Molin, "Relative intensity noise and frequency noise of a compact Brillouin laser made of $As_{38}Se_{62}$ suspended-core chalcogenide fiber" Optics Lett. **37**, 1157 (2012).

[8] H. Lee, T. Chen, Jiang Li, K. Y. Yang, S. Jeon, O. Painter and K. J. Vahala, "Chemically etched ultrahigh-Q wedge-resonator on a silicon chip," Nat. Photon. **6**, 369 (2012).

[9] R. Pant, C. G. Poulton, D.-Y. Choi, H. Mcfarlane, S. Hile, E. Li, L. Thevenaz, B. Luther-Davies, S. J. Madden, and B. J. Eggleton, "On-chip stimulated Brillouin Scattering," Opt. Express **19**, 8285 (2011).

[10] R. Pant, A. Byrnes, C. G. Poulton, E. Li, D.-Y. Choi, S. Madden, B. Luther-Davies, and B. J. Eggleton, "Photonic chip-based tunable slow and fast light via stimulated Brillouin scattering," Optics Lett. **37**, 969 (2012).

[11] S. Levy, V. Lyubin, M. Klebanov, J. Scheuer and A. Zadok, "Stimulated Brillouin scattering amplification in centimeter-long directly written chalcogenide waveguides," Optics Lett. **37**, 5112 (2012).

[12] R. Pant, E. Li, D.-Y. Choi, C. G. Poulton, S. J. Madden, B. Luther-Davies, and B. J. Eggleton, "Cavity enhanced stimulated Brillouin scattering in an optical chip for multiorder Stokes generation," Opt. Lett. **36**, 3687-3689 (2011).

[13] S. Madden, Z. Jin, D. Choi, S. Debbarma, D. Bulla, and B. Luther-Davies, "Low loss coupling to sub-micron thick rib and nanowire waveguides by vertical tapering," Opt. Express **21**, 3582 (2013).

[14] P. T. Rakich, C. Reinke, R. Camacho, P. Davids, and Z. Wang, "Giant Enhancement of Stimulated Brillouin Scattering in the Subwavelength Limit" Physical Review X **2**, 011008 (2012).

[15] S. Norcia, S. Tonda-Goldstein, D. Dolfi, J.-P. Huignard, and R. Frey, "Efficient single-mode Brillouin fiber laser for low-noise optical carrier reduction of microwave signals," Optics Lett. **28**, 1888 (2003).

[16] L. E. Richter, H. I. Mandelberg, M. S. Kruger, and P. A. McGrath, "Linewidth determination from self-heterodyne measurement with subcoherence delay times," IEEE J. Q. Electron. **22**, 2070 (1986).

[17] H. G. Winful, I. V. Kabakova, and B. J. Eggleton, "Model for distributed feedback Brillouin lasers" Opt. Express **21**, 16191 (2013).